\documentclass[journal,twoside,web]{ieeecolor}
\usepackage{tmi}
\usepackage{cite}
\usepackage{amsmath,amssymb,amsfonts}
\usepackage{algorithmic}
\usepackage{graphicx}
\usepackage{textcomp}
\usepackage{mathtools}
\usepackage{algorithm}
\usepackage{algorithmic}
\usepackage{textcmds}
\usepackage{url}
\usepackage{booktabs}
\usepackage{multirow} 

\definecolor{revisedColor}{rgb}{0,0,1}

\def\BibTeX{{\rm B\kern-.05em{\sc i\kern-.025em b}\kern-.08em
    T\kern-.1667em\lower.7ex\hbox{E}\kern-.125emX}}
\begin{document}
\title{Bayesian Kinetic Model Selection in Dynamic PET via Consistency Models}
\author{Yun Zhao, Qinlin Gu, Andrew J. Reader, Yanan Fan, Georgios I. Angelis, Katya Amadita, Gaelle Emvalomenos, Evan D. Morris, Guobao Wang, Steven R. Meikle
\thanks{This work is supported by the Australian Research Council (Discovery Projects grant DP230102070).}
\thanks{Yun Zhao (yun.zhao@sydney.edu.au) and Katya Amadita are with the School of Health Sciences and the Brain and Mind Centre, the University of Sydney, Australia.}
\thanks{Qinlin Gu is with the School of Mathematics and Statistics and the Brain and Mind Centre, the University of Sydney, Australia.}
\thanks{Andrew J. Reader is with the School of Biomedical Engineering and Imaging Sciences, King’s College London, UK.}
\thanks{Yanan Fan is with CSIRO Technology, Australia.}
\thanks{Georgios I. Angelis is with the School of Biomedical Engineering and Sydney Imaging, the University of Sydney, Australia.}
\thanks{Gaelle Emvalomenos and Evan D. Morris are with the Yale School of Medicine, USA.}
\thanks{Guobao Wang is with the Department of Radiology, University of California Davis Health, USA.}
\thanks{Steven R. Meikle is with the School of Health Sciences, Sydney Imaging and the Brain and Mind Centre, the University of Sydney, Australia.}
}

\maketitle

\begin{abstract}
Accurate voxel-wise parametric imaging in dynamic positron emission tomography (PET), particularly with total-body PET (TB-PET) scanners, requires selecting the appropriate kinetic model for diverse organs and lesions. 
Conventional nonlinear least squares fitting paired with the Akaike or Bayesian Information Criterion (NLS+AIC/BIC) is computationally slow, sensitive to local minima, and prone to selecting simple models under high voxel-level noise, whereas classical Bayesian methods such as Markov chain Monte Carlo (MCMC) and Approximate Bayesian Computation (PET-ABC) are too slow for whole-body imaging. In this work, we propose a consistency model (CM) that performs simultaneous Bayesian kinetic model selection and parameter posterior estimation. 
By combining the continuous kinetic parameters and discrete model index into one target vector conditioned on the measured time-activity curve (TAC) and input function, the CM samples from the joint posterior distribution $p(\boldsymbol{\theta}, m | \boldsymbol{y})$ in seconds per slice with a GPU workstation of 4 NVIDIA A100. 
We validated the framework on two simulated and clinical applications: whole-body $[\text{}^{18}\text{F}]\text{FDG}$ TB-PET across seven compartment models, and brain $[\text{}^{11}\text{C}]\text{raclopride}$ PET for detecting smoking-induced dopamine release between two neurotransmitter models. 
In simulations, the CM achieved 97.97\% model selection accuracy in the $[\text{}^{18}\text{F}]\text{FDG}$ application (vs.\ 58.67\% for NLS+AIC) and 91.4\% accuracy in the $[\text{}^{11}\text{C}]\text{raclopride}$ application (vs.\ 73.5\% for NLS+BIC), while closely matching reference MCMC parameter posteriors. 
In clinical applications, CM generated whole-body FDG model selection maps, in which each regions was assigned literature established sensible kinetic models; CM also successfully mapped voxels in the brain where dopamine release was induced by smoking.
\end{abstract}


\section{Introduction}
Dynamic positron emission tomography (PET) combined with kinetic modeling goes beyond static standardized uptake value (SUV) imaging by quantifying in vivo physiological processes such as blood flow, tissue perfusion, receptor binding, and glucose metabolism~\cite{phelps1979tomographic, ichise2003linearized}. 
With the rapid clinical and research adoption of long-axial-field-of-view and total-body PET (TB-PET) scanners worldwide, it is now possible to image all organs and lesions simultaneously with high sensitivity and temporal resolution while extracting an image-derived arterial input function directly from major vessels~\cite{liu2021kinetic, sari2022first, dias2022normal}. 
However, generating accurate whole-body parametric maps requires choosing the right kinetic model at each voxel. For FDG, different organs and lesions exhibit distinct vascular supplies and tracer kinetics, ranging from simple blood-to-tissue exchange in the lungs to irreversible phosphorylation in the brain and tumors~\cite{phelps1979tomographic}, multi-compartmental kinetics in skeletal muscle~\cite{bertoldo2001kinetic}, and dual hepatic arterial and portal venous blood supply in the liver~\cite{wang2018dynamic, wang2022total}. Even within a single organ, pathology alters model structure. For example, liver metastases are supplied mainly by the hepatic artery rather than the dual hepatic circulation of surrounding healthy parenchyma~\cite{wang2022total}. Similarly, in dynamic brain PET, detecting transient stimulus-induced neurotransmitter release requires distinguishing a time-varying neurotransmitter response model from a steady state receptor model at each voxel~\cite{normandin2012linear, morris2024modeling}. 
Applying a single fixed kinetic model across all voxels ignores this physiological diversity and introduces bias into kinetic parameter estimates~\cite{wang2018dynamic, wang2022total}.

Conventionally, voxel-wise kinetic model selection is performed by fitting every candidate model separately to each time-activity curve (TAC) using nonlinear least squares (NLS) and selecting the model that minimizes the Akaike Information Criterion (AIC) or Bayesian Information Criterion (BIC)~\cite{normandin2012linear, zuo2020multiparametric, wang2022total}. 
Despite its widespread use, this frequentist approach has three major limitations in voxel-level PET imaging. 
First, iteratively fitting a library of compartment models across millions of TB-PET voxels is computationally expensive and prone to local minima. 
Second, AIC and BIC rely on asymptotic Gaussian noise assumptions and apply fixed penalties based on the number of free parameters. 
Whereas existing research has showed that the noise properties derived from the expectation-maximization (EM) reconstruction algorithm follow a log-normal distribution\cite{barrett1994noise,wilson1994noise}.
Under high voxel-level noise, random fluctuations mask the subtle curve-fitting improvements of additional tissue compartments, dual blood inputs, or brief neurotransmitter responses, causing AIC and BIC to over-penalize complex models~\cite{wang2022total, liu2019controlling, morris2024modeling}. 
Third, information criteria only output a single binary or categorical ``winning'' model without quantifying model selection uncertainty, even in heterogeneous boundary voxels or low-count tissues where multiple models are plausible.

Bayesian inference overcomes these limitations by computing full posterior distributions $p(\boldsymbol{\theta} | m, \mathbf{y})$ and marginal posterior model probabilities $P(m | \mathbf{y})$ directly from the noisy measurement data~\cite{gelman1995bayesian}. 
In dynamic PET, classical Bayesian methods such as Markov chain Monte Carlo (MCMC) paired with leave-one-out cross-validation (LOO-CV)~\cite{irace2020bayesian, hoffman2014no, vehtari2017practical} and likelihood-free Approximate Bayesian Computation (PET-ABC)~\cite{fan2021pet} are useful methods for uncertainty quantification and model selection. 
For MCMC+LOO, computation requires thousands of sequential leapfrog steps for every candidate model at each voxel. PET-ABC relies on choosing appropriate summary statistics and calibrating acceptance thresholds across millions of forward simulations. 
These computational challenges motivate faster approaches to voxel-wise Bayesian model selection and parameter estimation from dynamic PET.

Recent advances in score-based diffusion models~\cite{ho2020denoising, song2020score} and consistency models (CMs)~\cite{song2023consistency} offer a fast alternative for Bayesian inference by learning to map Gaussian noise directly to complex posterior distributions in only a few network evaluations. 
Recently, we introduced a CM framework for voxel-wise kinetic posterior estimation with two-tissue compartment models in dynamic TB-PET, achieving MCMC-level accuracy while accelerating inference by over five orders of magnitude~\cite{zhao2026generative}. 
In this study, we augment this framework to perform simultaneous Bayesian kinetic model selection and parameter estimation across a candidate model space. By combining the continuous kinetic parameter vector and the discrete model index into the target vector, our CM learns the joint posterior distribution $p(\boldsymbol{\theta}, m | \mathbf{y})$ conditioned on the measured TAC and input function. 
During inference, drawing samples from the CM yields voxel-wise posterior model probabilities and model-specific parameter posteriors in seconds per slice. 
We validate our approach in both comprehensive simulations and clinical scans across two applications: whole-body $[\text{}^{18}\text{F}]\text{FDG}$ TB-PET model selection across seven compartment models, and brain $[\text{}^{11}\text{C}]\text{raclopride}$ PET activation mapping to detect transient smoking-induced dopamine release.

\section{Methods}\label{methods}
\subsection{Consistency Models}
Consistency models (CMs)~\cite{song2023consistency} accelerate sampling by replacing the hundreds or thousands of reverse diffusion steps typically used by score-based diffusion models~\cite{ho2020denoising,song2020score} with as few as one forward pass.
Along a continuous-time probability flow ordinary differential equation (PF-ODE) trajectory that perturbs data $\mathbf{x}_0 \sim p_{\text{data}}(\mathbf{x})$ toward a Gaussian prior over noise scales $t \in [\epsilon, T]$ ($\epsilon > 0$), a CM learns a consistency function $f_{\phi}(\mathbf{x}_t, t)$ mapping any noisy state $\mathbf{x}_t$ directly back to its trajectory origin $\mathbf{x}_{\epsilon}$. This mapping satisfies the self-consistency property $f_{\phi}(\mathbf{x}_{t_1}, t_1) = f_{\phi}(\mathbf{x}_{t_2}, t_2) = \mathbf{x}_{\epsilon}, \quad \forall t_1, t_2 \in [\epsilon, T]$, with boundary condition $f_{\phi}(\mathbf{x}_{\epsilon}, \epsilon) = \mathbf{x}_{\epsilon}$, which is enforced analytically via skip connections, $f_{\phi}(\mathbf{x}_t, t) = c_{\text{skip}}(t)\,\mathbf{x}_t + c_{\text{out}}(t)\,F_{\phi}(\mathbf{x}_t, t)$, where $F_{\phi}(\cdot)$ is a deep neural network and $c_{\text{skip}}(\epsilon) = 1$, $c_{\text{out}}(\epsilon) = 0$~\cite{song2023consistency}.

Building on this, we recently introduced a conditional CM framework for voxel-wise Bayesian estimation of kinetic parameters in dynamic total-body PET~\cite{zhao2026generative}. The framework uses a conditional mapping $f_{\phi}(\boldsymbol{\theta}_t, t, \mathbf{y})$, where $\boldsymbol{\theta}_t$ is the noisy parameter vector at diffusion time $t$ and $\boldsymbol{\theta}_0 \in \mathbb{R}^{d}$ is the ground-truth kinetic parameter vector. The conditioning vector $\mathbf{y} \in \mathbb{R}^{2M}$ concatenates the $M$-frame voxel TAC with the arterial input function (AIF). After training on synthetic pairs $(\boldsymbol{\theta}_0, \mathbf{y})$, the network takes $\mathbf{y}$ as input and maps independent Gaussian noise draws directly to samples from the target posterior $p(\boldsymbol{\theta} | \mathbf{y})$. The framework achieved accuracy comparable to that of MCMC while accelerating voxel-wise inference by more than five orders of magnitude~\cite{zhao2026generative}.

In this study, we extend this framework to perform simultaneous Bayesian model selection and parameter estimation across a candidate model space $\mathcal{M} = \{M_1, \dots, M_K\}$. We augment the kinetic parameter vector $\boldsymbol{\theta}$ with the discrete model index $m \in \{1, \dots, K\}$ to form a target vector $\mathbf{x} = [\boldsymbol{\theta}^{\top}, m]^{\top}$, enabling $f_{\phi}(\mathbf{x}_t, t, \mathbf{y})$ to map Gaussian noise directly to the joint posterior $p(\boldsymbol{\theta}, m | \mathbf{y})$. 
At inference, the learned $f_{\phi}(\mathbf{x}_t, t, \mathbf{y})$ conditions on the measured TAC and input function $\mathbf{y}$ to transform $S$ Gaussian noise draws $\mathbf{z}^{(s)} \sim \mathcal{N}(\mathbf{0}, \mathbf{I})$ into joint posterior samples $\{(\hat{\boldsymbol{\theta}}_{0}^{(s)}, \hat{m}_{0}^{(s)})\}_{s=1}^{S} \sim p(\boldsymbol{\theta}, m | \mathbf{y})$ in a single pass. 
Marginal posterior model probabilities $P(m = k | \mathbf{y})$ are then obtained directly from the sample proportions assigned to each model $k$, while the corresponding sample subsets characterize the conditional parameter posteriors $p(\boldsymbol{\theta} | m = k, \mathbf{y})$.
For network architecture, training, and sampling, we adopt the conditional 1D U-Net backbone and standalone consistency training algorithm detailed in our previous work~\cite{zhao2026generative}. Briefly, the 1D U-Net comprises four encoder and decoder stages (channel progression $32 \to 64 \to 128 \to 256$) with residual skip connections, conditioned on $\mathbf{y}$ via dense projections and on $t$ via Gaussian Fourier feature embeddings combined through SiLU activations. During training, the student network parameters are optimized using AdamW to match an exponential moving average (EMA) teacher network (fixed decay rate $\mu = 0.999$) across adjacent noise levels drawn from a discretized schedule of $N_k = 15$ steps. At inference, posterior samples are generated using 3-step multistep consistency sampling; full algorithm and architectural specifications are provided in~\cite{zhao2026generative}.

\subsection{Application of CMs to Model Selection}
\subsubsection{Compartment Models}
In the first case study, we consider seven compartment models ($m \in \{0, \dots, 6\}$) for voxel-wise model selection in dynamic FDG TB-PET. Let $C_p(t)$ denote the arterial input function (AIF) and $t_d$ the tissue arrival delay. Single-input models ($m \in \{0, 1, 3, 5\}$) use the delayed arterial input, $C_{\text{in}}(t) = C_p(t - t_d)$. Dual-input models ($m \in \{2, 4, 6\}$) represent two contributions to the tissue blood supply. The first is the delayed arterial input, and the second, $C_{disp}(t)$, is obtained by passing that same input through a first-order transit model, which introduces additional lag and dispersion~\cite{wang2018dynamic}. Here, $k_a$ ($\mathrm{min}^{-1}$) is the transit rate and $f_A$ is the fraction of the total input contributed by the direct arterial supply. The two components are combined as follows.
\begin{align}
    \frac{dC_{disp}(t)}{dt} &= k_a \bigl(C_p(t - t_d) - C_{disp}(t)\bigr), \label{eq:cpv}\\
    C_{\text{in}}(t) &= f_A C_p(t - t_d) + (1 - f_A) C_{disp}(t). \label{eq:cin}
\end{align}
The total measured tissue activity is the sum of the tissue and blood contributions, weighted by the fractional blood volume $V_b$.
\begin{equation}
    C_{\text{total}}(t) = (1 - V_b)\,C_{\text{tissue}}(t) + V_b\,C_{\text{in}}(t). \label{eq:ctotal}
\end{equation}

The seven models comprise a blood-pool model and single- and dual-input variants of the one-, two-, and three-tissue compartment models. Each dual-input variant uses the input in~\eqref{eq:cin} and adds $k_a$ and $f_A$, in that order, immediately before $t_d$ in the corresponding parameter vector below. Throughout, $K_1$ is the influx rate ($\mathrm{mL}\,\mathrm{cm}^{-3}\,\mathrm{min}^{-1}$), and $k_2$ is the efflux rate. All rate constants $k_2$ through $k_6$ are in $\mathrm{min}^{-1}$.

\textbf{0TCM ($m=0$).} For blood-pool voxels, the model assumes no extravascular uptake, giving $C_{\text{total}}(t)=V_b C_p(t-t_d)$ and $\boldsymbol{\theta}=[V_b,t_d]^{\top}$.

\textbf{1TCM and 1TCM-Dual ($m=1,2$).} A single tissue compartment exchanges tracer with plasma without phosphorylation, representing rapid equilibration or negligible trapping.
\begin{equation}
    \frac{dC_T(t)}{dt}=K_1 C_{\text{in}}(t)-k_2 C_T(t). \label{eq:1tcm}
\end{equation}
Here, $C_{\text{tissue}}(t)=C_T(t)$ and the single-input parameter vector is $\boldsymbol{\theta}=[K_1,k_2,V_b,t_d]^{\top}$.

\textbf{2TCM and 2TCM-Dual ($m=3,4$), and 3TCM and 3TCM-Dual ($m=5,6$).} The standard 2TCM~\cite{phelps1979tomographic} models free ($C_f$) and phosphorylated ($C_m$) tracer, with phosphorylation and dephosphorylation rates $k_3$ and $k_4$, respectively. The 3TCM adds a reversible compartment ($C_{ns}$), with entry and exit rates $k_5$ and $k_6$, to capture multiexponential clearance or nonspecific binding. The full system is
\begin{align}
    \frac{dC_f(t)}{dt} &= K_1 C_{\text{in}}(t) - (k_2+k_3+k_5)C_f(t) \nonumber\\
                       &\quad + k_4 C_m(t) + k_6 C_{ns}(t), \label{eq:3tcm_cf}\\
    \frac{dC_m(t)}{dt} &= k_3 C_f(t) - k_4 C_m(t), \label{eq:3tcm_cm}\\
    \frac{dC_{ns}(t)}{dt} &= k_5 C_f(t) - k_6 C_{ns}(t). \label{eq:3tcm_cns}
\end{align}
The 3TCM has $C_{\text{tissue}}(t)=C_f(t)+C_m(t)+C_{ns}(t)$ and $\boldsymbol{\theta}=[K_1,k_2,k_3,k_4,k_5,k_6,V_b,t_d]^{\top}$. The 2TCM omits the $C_{ns}$ equation and the terms involving $k_5$ and $k_6$, giving $C_{\text{tissue}}(t)=C_f(t)+C_m(t)$ and $\boldsymbol{\theta}=[K_1,k_2,k_3,k_4,V_b,t_d]^{\top}$.

\subsubsection{Neurotransmitter Models}
In the second case study, we detect transient task- or stimulus-induced dopamine release at the voxel level in brain PET with [$^{11}\mathrm{C}$]raclopride. Both models use the TAC $C_r(t)$ from a receptor-devoid reference region (e.g., cerebellum). The target tissue TAC $C_t(t)$ follows
\begin{align}
    C_t(t) &= R_1 C_r(t) + k_2 \int_0^t C_r(u)\,\mathrm{d}u \nonumber\\
           &\quad - k_{2a} \int_0^t C_t(u)\,\mathrm{d}u - \gamma B(t).
    \label{eq:lpntpet}
\end{align}
Here, $R_1=K_1/K_1'$ is the target-to-reference tracer delivery ratio, and $k_2$ and $k_{2a}$ (both in $\mathrm{min}^{-1}$) are the target tissue efflux and baseline apparent efflux rate constants, respectively.

\textbf{Multilinear reference tissue model (MRTM, $m=0$).} The null model describes baseline kinetics by setting $\gamma=0$~\cite{ichise2003linearized}, with $\boldsymbol{\theta}_{\text{MRTM}}=[R_1,k_2,k_{2a}]^{\top}$.

\textbf{Linear parametric neurotransmitter PET (lp-ntPET, $m=1$).} The term $\gamma B(t)$ accounts for dopamine-induced tracer displacement~\cite{normandin2012linear}, where $\gamma$ ($\mathrm{min}^{-1}$) is the activation amplitude and
\begin{align}
    B(t) &= \int_0^t C_t(u)\,g(u)\,\mathrm{d}u, \label{eq:bt}\\
    g(t) &= \tau(t)^\alpha \exp\!\bigl(\alpha[1-\tau(t)]\bigr),
            \quad t\ge t_D, \label{eq:ht}\\
    \tau(t) &= \frac{t-t_D}{t_P-t_D}. \label{eq:tau}
\end{align}
The gamma-variate response is zero for $t<t_D$. The onset $t_D$ and peak $t_P>t_D$ are measured in minutes from scan start, and $\alpha$ controls response sharpness. The parameter vector is $\boldsymbol{\theta}_{\text{lp-ntPET}}=[R_1,k_2,k_{2a},\gamma,t_D,t_P,\alpha]^{\top}$.

\subsubsection{Simulation and Training Details}
In both case studies, we sampled kinetic parameters from the model-specific priors in Table~\ref{tab:priors}, informed by the PET literature~\cite{phelps1979tomographic,liu2021kinetic,sari2022first,dias2022normal,wang2018dynamic,zuo2020multiparametric,bertoldo2001kinetic,wang2022total,normandin2012linear,irace2020bayesian,fan2021pet}. For the FDG study, we simulated TACs using Feng-model arterial input functions~\cite{feng1993models}, with parameter ranges derived from image-derived input functions in our clinical [$^{18}\mathrm{F}$]FDG cohort.

To approximate the nonnegative, right-skewed variability of expectation-maximization (EM) reconstructions, we applied mean-preserving log-normal noise~\cite{barrett1994noise,wilson1994noise}. With independent $\epsilon_i\sim\mathcal{N}(0,1)$, the noisy TAC is
\begin{equation}
    \widetilde{C}_{\text{total}}(t_i)
    = \bar{C}_{\text{total}}(t_i)
      \exp\!\left(\eta_i\epsilon_i-\tfrac{1}{2}\eta_i^2\right),
    \label{eq:measurement_noise}
\end{equation}
where $\bar{C}_{\text{total}}(t_i)=\max\{C_{\text{total}}(t_i),10^{-8}\}$ ensures strict positivity and numerical stability. The log-space and frame standard deviations are
\begin{equation}
    \eta_i=\sqrt{\ln\!\left(1+\frac{\sigma_i^2}{\bar{C}_{\text{total}}(t_i)^2}\right)},
    \quad
    \sigma_i=\ell\sqrt{\frac{\bar{C}_{\text{total}}(t_i)e^{\lambda t_i}}{\Delta t_i}},
    \label{eq:noise_sd}
\end{equation}
where $\Delta t_i$ is the frame duration and $\lambda=\ln 2/T_{1/2}$, with $T_{1/2}=109.8$~min for $^{18}\mathrm{F}$ and $20.33$~min for $^{11}\mathrm{C}$. We sampled $\ell\sim\mathcal{U}(3,7)$ for [$^{18}\mathrm{F}$]FDG and $\ell\sim\mathcal{U}(0,4)$ for [$^{11}\mathrm{C}$]raclopride.

For each candidate model, we generated $2\times10^6$ training, $10^5$ validation, and $1{,}000$ test pairs. We trained the CM for up to $5{,}000$ epochs using AdamW with an initial learning rate of $2\times10^{-3}$ and a plateau scheduler (factor $0.75$, patience $25$ epochs, minimum learning rate $10^{-6}$). We used $N_k=15$, $\sigma_{\max}=1.0$, and an EMA decay of $\mu=0.999$, and evaluated the EMA teacher checkpoint with the lowest validation loss on the test set.

\begin{table*}[!t]
\caption{\textsc{Uniform Prior Distributions $\mathcal{U}(\min, \max)$ for Kinetic Parameters.}}
\label{tab:priors}
\centering
\scriptsize
\setlength{\tabcolsep}{1.2pt}
\renewcommand{\arraystretch}{1.15}

\begin{tabular*}{\textwidth}{@{\extracolsep{\fill}} l c c c c c c c c c c}
\toprule
\multicolumn{11}{l}{\textbf{A. Compartment models}} \\
Model & $K_1$ & $k_2$ & $k_3$ & $k_4$ & $k_5$ & $k_6$ & $V_b$ & $k_a$ & $f_A$ & $t_d$ (s) \\
\midrule
0TC ($m=0$)   & --          & --          & --          & --          & --          & --          & $(0, 1.00)$ & --           & --          & $(-10, 50)$ \\
1TC ($m=1$)   & $(0, 2.00)$ & $(0, 2.00)$ & --          & --          & --          & --          & $(0, 1.00)$ & --           & --          & $(-10, 50)$ \\
1TC-D ($m=2$) & $(0, 2.00)$ & $(0, 2.00)$ & --          & --          & --          & --          & $(0, 1.00)$ & $(0, 10.00)$ & $(0, 1.00)$ & $(-10, 50)$ \\
2TC ($m=3$)   & $(0, 2.00)$ & $(0, 2.00)$ & $(0, 0.50)$ & $(0, 0.20)$ & --          & --          & $(0, 1.00)$ & --           & --          & $(-10, 50)$ \\
2TC-D ($m=4$) & $(0, 2.00)$ & $(0, 2.00)$ & $(0, 0.50)$ & $(0, 0.20)$ & --          & --          & $(0, 1.00)$ & $(0, 10.00)$ & $(0, 1.00)$ & $(-10, 50)$ \\
3TC ($m=5$)   & $(0, 2.00)$ & $(0, 2.00)$ & $(0, 0.50)$ & $(0, 0.20)$ & $(0, 1.00)$ & $(0, 0.50)$ & $(0, 1.00)$ & --           & --          & $(-10, 50)$ \\
3TC-D ($m=6$) & $(0, 2.00)$ & $(0, 2.00)$ & $(0, 0.50)$ & $(0, 0.20)$ & $(0, 1.00)$ & $(0, 0.50)$ & $(0, 1.00)$ & $(0, 10.00)$ & $(0, 1.00)$ & $(-10, 50)$ \\
\midrule
\multicolumn{11}{l}{\textbf{B. Neurotransmitter models}} \\
Model & $R_1$ & $k_2$ & $k_{2a}$ & $\gamma$ & $t_D$ (min) & $t_P$ (min) & $\alpha$ & \multicolumn{3}{c}{} \\
\midrule
MRTM ($m=0$)     & $(0.40, 1.60)$ & $(0, 0.60)$ & $(0, 0.20)$ & --          & --         & --           & --          & \multicolumn{3}{c}{} \\
lp-ntPET ($m=1$) & $(0.40, 1.60)$ & $(0, 0.60)$ & $(0, 0.20)$ & $(0, 0.20)$ & $(30, 40)$ & $(t_D+3, 60)$ & $(0, 4.00)$ & \multicolumn{3}{c}{} \\
\bottomrule
\end{tabular*}

\par\vspace{1pt}
\begin{minipage}{\textwidth}
\footnotesize
\raggedright
\textit{Units.} $K_1$ is in $\mathrm{mL}\cdot\mathrm{cm}^{-3}\cdot\mathrm{min}^{-1}$.
The rate constants $k_2, k_3, k_4, k_5, k_6, k_a, k_{2a}, \gamma$
are in $\mathrm{min}^{-1}$. $V_b, f_A, R_1, \alpha$ are unitless.
$t_d$ is in seconds, and $t_D, t_P$ are in minutes.
\end{minipage}
\end{table*}

\subsection{Performance Evaluation}
\subsubsection{Reference Methods}
For voxel-wise model selection, we fitted each candidate model $m\in\mathcal{M}$ by NLS from five starting points to mitigate local minima. These comprised the prior midpoint and four uniform draws within the parameter bounds. 
We retained the converged estimate $\hat{\boldsymbol{\theta}}_m$ with the smallest residual sum of squares across the $M$ frames, $\mathrm{RSS}_m=\sum_{i=1}^{M}[y(t_i)-\hat{C}_{\text{total}}(t_i;\hat{\boldsymbol{\theta}}_m)]^2$. 
We then selected $\hat{m}=\arg\min_{m\in\mathcal{M}}\mathrm{IC}(m)$ using the corrected Akaike information criterion ($\mathrm{AIC}_c$)~\cite{wang2022total} or the Bayesian information criterion (BIC)~\cite{normandin2012linear},
\begin{align}
    \mathrm{AIC}_c(m) &= M\ln\!\left(\frac{\mathrm{RSS}_m}{M}\right)
    +2n_m+\frac{2n_m(n_m+1)}{M-n_m-1}, \label{eq:aicc}\\
    \mathrm{BIC}(m) &= M\ln\!\left(\frac{\mathrm{RSS}_m}{M}\right)
    +n_m\ln M, \label{eq:bic}
\end{align}
where $n_m$ is the number of free parameters in model $m$.

To approximate the reference posterior, we used MCMC~\cite{gelman1995bayesian} with the No-U-Turn Sampler (NUTS)~\cite{hoffman2014no} in NumPyro~\cite{phan2019composable}. For each TAC, NUTS used a target acceptance probability of $0.95$ and $2{,}000$ warm-up iterations for step-size and mass-matrix adaptation, followed by $1{,}000$ posterior draws.
For Bayesian model selection in the [$^{11}\mathrm{C}$]raclopride study, we fitted MRTM and lp-ntPET separately by MCMC and compared them using Pareto-smoothed importance sampling leave-one-out cross-validation (PSIS-LOO-CV)~\cite{vehtari2017practical}. PSIS-LOO-CV estimates the expected log pointwise predictive density ($\mathrm{elpd}_{\mathrm{loo}}$) from posterior pointwise log-likelihood draws without refitting. We selected the model with the higher $\mathrm{elpd}_{\mathrm{loo}}$, favoring out-of-sample predictive accuracy and accounting for overfitting.

\subsubsection{Clinical Dataset}
For the [\(^{18}\text{F}\)]FDG TB-PET case study, we analyzed a clinical cohort of 13 oncological subjects who underwent 50-min dynamic scans on a Siemens Biograph Vision Quadra scanner (Siemens Healthineers, Knoxville, USA) following intravenous bolus injection of [\(^{18}\text{F}\)]FDG ($146.67 \pm 11.50$~MBq), approved by the Northern Sydney Local Health District Human Research Ethics Committee with written informed consent~\cite{zhao2026generative}. List-mode data were binned into 35 dynamic frames ($220 \times 220 \times 645$ matrix) and reconstructed using time-of-flight ordinary Poisson ordered-subsets expectation maximization (OP-OSEM; 8 iterations, 5 subsets, maximum ring difference 322) with standard physical corrections (normalization, randoms, attenuation, scatter, and decay), followed by spatiotemporal HYPR-LR denoising~\cite{christian2010dynamic}. For each of the 13 subjects, an image-derived AIF was extracted from the ascending aorta by segmenting the native CT image with MOOSE~\cite{sundar2022fully} and morphologically eroding the mask by a two-voxel radius to mitigate partial volume effects~\cite{de2006comparison}; this cohort of extracted AIFs was also used to drive the forward simulations during CM training. 
To demonstrate clinical voxel-wise model selection and parameter estimation, we focus primarily on a representative subject P001, a 54-year-old male diagnosed with a pancreatic neuroendocrine tumor and a known hepatic metastasis at the inferior right margin of the liver.

For the [\(^{11}\text{C}\)]raclopride case study, we analyzed a 90-min dynamic brain scan acquired on a High Resolution Research Tomograph (HRRT; Siemens Healthineers, Knoxville, USA) from a human smoker following 1~week under a placebo patch and overnight abstinence (approved by the Yale Human Investigation and Radiation Safety Committees)~\cite{zakiniaeiz2022nicotine}. Beginning at 35~min post-injection, the subject smoked a cigarette inside the scanner for approximately 3~min. Reconstructed dynamic frames were denoised with a $3 \times 3 \times 3$ voxel HYPR filter~\cite{christian2010dynamic} and spatially normalized to MNI template space, where a cerebellar reference-region TAC $C_r(t)$ was extracted and a 1,004-voxel mask delineating the precommissural striatum (ventral striatum, dorsal caudate, and dorsal putamen)~\cite{martinez2003imaging} was applied for voxel-wise model selection.

\begin{figure*}[t!]
    \centering
    \includegraphics[width=0.90\textwidth]{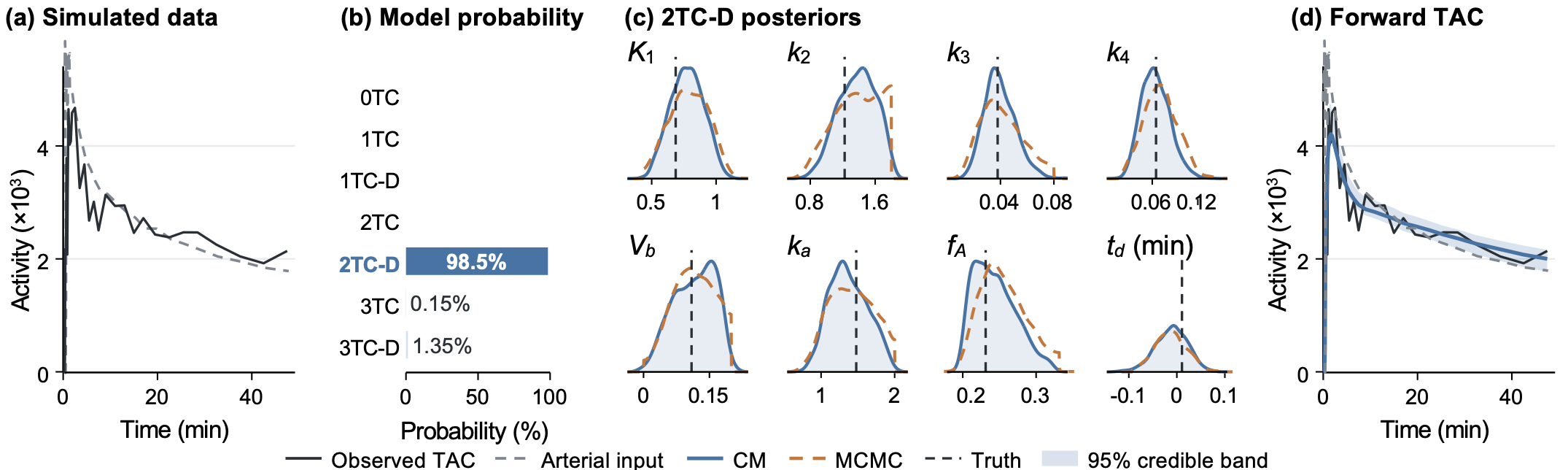}
    \caption{Example CM outputs for a simulated 2TC-D TAC: (a) Simulated TAC and arterial input; (b) Posterior model probabilities; (c) Parameter posteriors predicted by CM and MCMC; (d) Forward simulations from CM posterior samples. The blue line shows the mean simulated TAC, and the shaded band indicates the pointwise 95\% credible interval.
    }
    \label{fig: CM_simulated_2TCD_example}
\end{figure*}

\begin{figure*}[t!]
    \centering
    \includegraphics[width=0.85\textwidth]{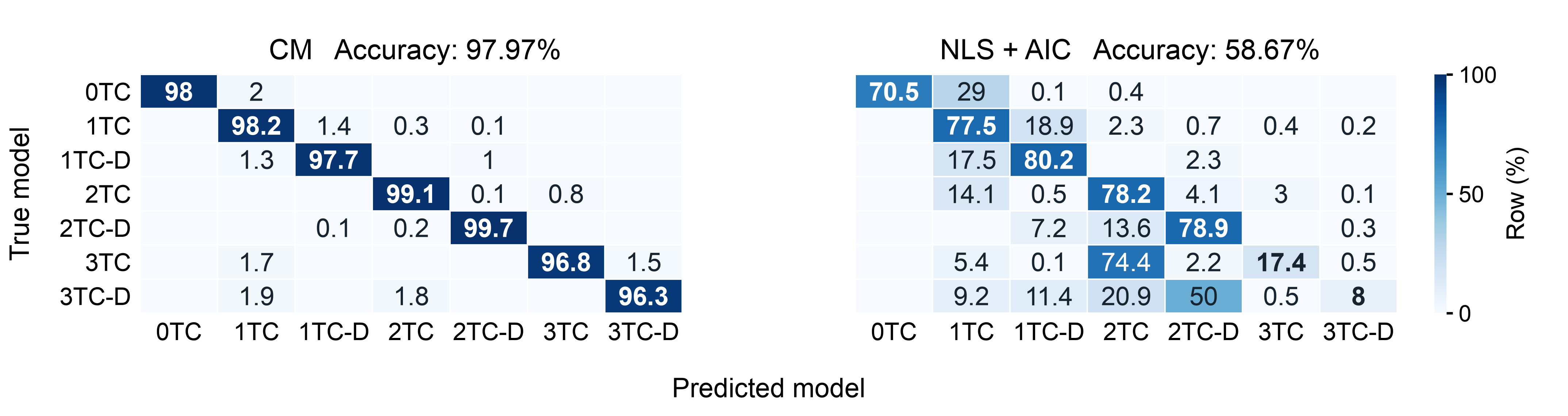}
    \caption{Confusion matrices for seven compartment model selection on simulated PET data using CM and NLS + AIC. Rows indicate true models and columns predicted models; entries show counts and row-normalized percentages. “D” denotes dual input.
    }
    \label{fig: compartment_confusion_matrix}
\end{figure*}

\subsubsection{Performance Metrics}
Model selection performance on the simulated test sets was evaluated using confusion matrices and overall classification accuracy, defined as $\text{Accuracy} = \frac{1}{N_{\text{test}}} \sum_{j=1}^{N_{\text{test}}} \mathbb{I}(\hat{m}_j = m_j^{*})$, where $m_j^{*}$ is the ground-truth model index, $\hat{m}_j$ is the predicted model index (given by $\arg\max_{k} P(m = k | \mathbf{y}_j)$ for CM or $\arg\min_{k} \text{IC}(k)$ for NLS), and $\mathbb{I}(\cdot)$ is the indicator function. To assess kinetic parameter posterior estimation for each true model, the distributional agreement between the CM posterior $q_{\text{CM}}(\theta | \mathbf{y})$ and the reference MCMC posterior $p_{\text{MCMC}}(\theta | \mathbf{y})$ (conditioned on the ground-truth model) was quantified using Kullback--Leibler divergence (KLD), computed as $D_{\text{KL}}(p_{\text{MCMC}} \parallel q_{\text{CM}}) = \frac{1}{S} \sum_{s=1}^{S} \ln \bigl(\hat{p}_{\text{MCMC}}(\theta^{(s)}) / \hat{q}_{\text{CM}}(\theta^{(s)})\bigr)$, where $\hat{p}_{\text{MCMC}}$ and $\hat{q}_{\text{CM}}$ denote kernel density estimates constructed from $S = 1,000$ posterior draws $\theta^{(s)} \sim p_{\text{MCMC}}(\theta | \mathbf{y})$. Point-estimate accuracy for each kinetic parameter was evaluated using the absolute percentage error (APE) between the CM posterior median $\hat{\theta}_{\text{med}}$ and the simulated ground-truth parameter $\theta^{*}$, defined as $\text{APE} = 100\% \times |\hat{\theta}_{\text{med}} - \theta^{*}| / |\theta^{*}|$.

\begin{table*}[!tb]
\centering
\caption{Parameter estimation performance. Each cell shows median KLD and APE (\%) over 1,000 simulated test TACs.}
\label{tab:posterior_performance}
\begingroup
\footnotesize
\setlength{\tabcolsep}{1.6pt}
\renewcommand{\arraystretch}{1.08}
\begin{tabular*}{\textwidth}{@{\extracolsep{\fill}}l*{10}{c}@{}}
\toprule
\multicolumn{11}{l}{\textbf{A. Compartment models case study}}\\
Model & $K_1$ & $k_2$ & $k_3$ & $k_4$ & $k_5$ & $k_6$ & $V_b$ & $k_a$ & $f_A$ & $t_d$\\
\midrule
0TC & -- & -- & -- & -- & -- & -- & $0.06\,/\,2.1$ & -- & -- & $0.14\,/\,8.4$\\
1TC & $0.11\,/\,3.2$ & $0.14\,/\,4.7$ & -- & -- & -- & -- & $0.18\,/\,6.1$ & -- & -- & $0.19\,/\,9.2$\\
1TC-D & $0.18\,/\,4.1$ & $0.24\,/\,5.8$ & -- & -- & -- & -- & $0.26\,/\,7.4$ & $0.33\,/\,6.9$ & $0.41\,/\,9.5$ & $0.24\,/\,10.1$\\
2TC & $0.15\,/\,3.8$ & $0.24\,/\,5.5$ & $0.36\,/\,8.7$ & $0.59\,/\,11.4$ & -- & -- & $0.24\,/\,7.1$ & -- & -- & $0.27\,/\,10.8$\\
2TC-D & $0.22\,/\,4.7$ & $0.32\,/\,6.6$ & $0.51\,/\,11.3$ & $0.71\,/\,12.9$ & -- & -- & $0.30\,/\,8.3$ & $0.41\,/\,8.1$ & $0.54\,/\,11.6$ & $0.32\,/\,10.2$\\
3TC & $0.25\,/\,5.1$ & $0.38\,/\,7.4$ & $0.58\,/\,12.9$ & $0.83\,/\,16.5$ & $0.69\,/\,15.6$ & $0.94\,/\,18.1$ & $0.35\,/\,9.2$ & -- & -- & $0.36\,/\,11.5$\\
3TC-D & $0.31\,/\,6.0$ & $0.47\,/\,8.8$ & $0.67\,/\,12.7$ & $0.95\,/\,18.3$ & $0.79\,/\,13.4$ & $0.98\,/\,18.6$ & $0.42\,/\,8.7$ & $0.52\,/\,8.9$ & $0.65\,/\,12.8$ & $0.44\,/\,10.1$\\
\bottomrule
\end{tabular*}

\vspace{3pt}
\begin{tabular*}{\textwidth}{@{\extracolsep{\fill}}l*{7}{c}@{}}
\toprule
\multicolumn{8}{l}{\textbf{B. Neurotransmitter models case study}}\\
Model & $R_1$ & $k_2$ & $k_{2a}$ & $\gamma$ & $t_D$ & $t_P$ & $\alpha$\\
\midrule
MRTM & $0.65\,/\,7.6$ & $0.78\,/\,13.9$ & $0.66\,/\,11.4$ & -- & -- & -- & --\\
lpnt-PET & $0.55\,/\,7.1$ & $1.32\,/\,15.2$ & $1.27\,/\,19.8$ & $0.82\,/\,15.8$ & $1.29\,/\,17.4$ & $0.86\,/\,11.2$ & $1.15\,/\,15.9$\\
\bottomrule
\end{tabular*}

\vspace{3pt}
\begin{minipage}{\textwidth}
\footnotesize
KLD is $D_{\mathrm{KL}}(p_{\mathrm{MCMC}}\Vert q_{\mathrm{CM}})$.
Posteriors predicted by MCMC are conditioned on the true model.
APE evaluates the CM posterior median against the simulated truth.
``--'': absent or fixed parameter.
\end{minipage}
\endgroup
\end{table*}

\section{Results}
\subsection{TB-PET FDG Model Selection}
Fig.\ref{fig: CM_simulated_2TCD_example} shows model selection and posterior estimation produced by the CM for an example simulated 2TC-D TAC. The CM assigns a posterior model probability of 98.5\% to the ground-truth 2TC-D model, with negligible probability mass allocated to competing three-tissue models (0.15\% for 3TC and 1.35\% for 3TC-D). Conditioned on the selected 2TC-D model, the marginal posterior distributions for all eight parameters closely match the reference MCMC posteriors and encapsulate the simulated ground-truth values within high-density regions, while forward simulations drawn from the CM posterior samples accurately recover the underlying noiseless TAC within a 95\% credible band.

Across all 7,000 simulated test TACs (1,000 per model), the CM achieves an overall classification accuracy of 97.97\%, substantially outperforming conventional NLS+AIC at 58.67\% (Fig.\ref{fig: compartment_confusion_matrix}). 
The CM achieves high accuracy in identifying the seven models, ranging from 96.3\% for 3TC-D up to 99.7\% for 2TC-D. 
In contrast, NLS+AIC struggles to identify the more complex three-tissue models under noise. While it achieves moderate accuracy for 0TC through 2TC-D (70.5\%–80.2\%), its accuracy drops to 17.4\% for 3TC and 8.0\% for 3TC-D, as most 3TC TACs are misclassified as 2TC (74.4\%) and most 3TC-D TACs are misclassified as 2TC-D (50.0\%) or 2TC (20.9\%).
Besides, Table~\ref{tab:posterior_performance} Part~A confirms that CM parameter posteriors closely match the reference MCMC distributions across all seven models. The primary rate constant $K_1$ and blood volume $V_b$ are recovered with median KLD values of 0.11--0.31 and 0.06--0.42, respectively. Even for other micro-parameters ($k_3$--$k_6$) and dual-input dispersion parameters ($k_a, f_A$), median KLD values remain below 0.98 with median APEs between 6.9\% and 18.6\%.

\begin{figure}[tb!]
    \centering
    \includegraphics[width=0.95\columnwidth]{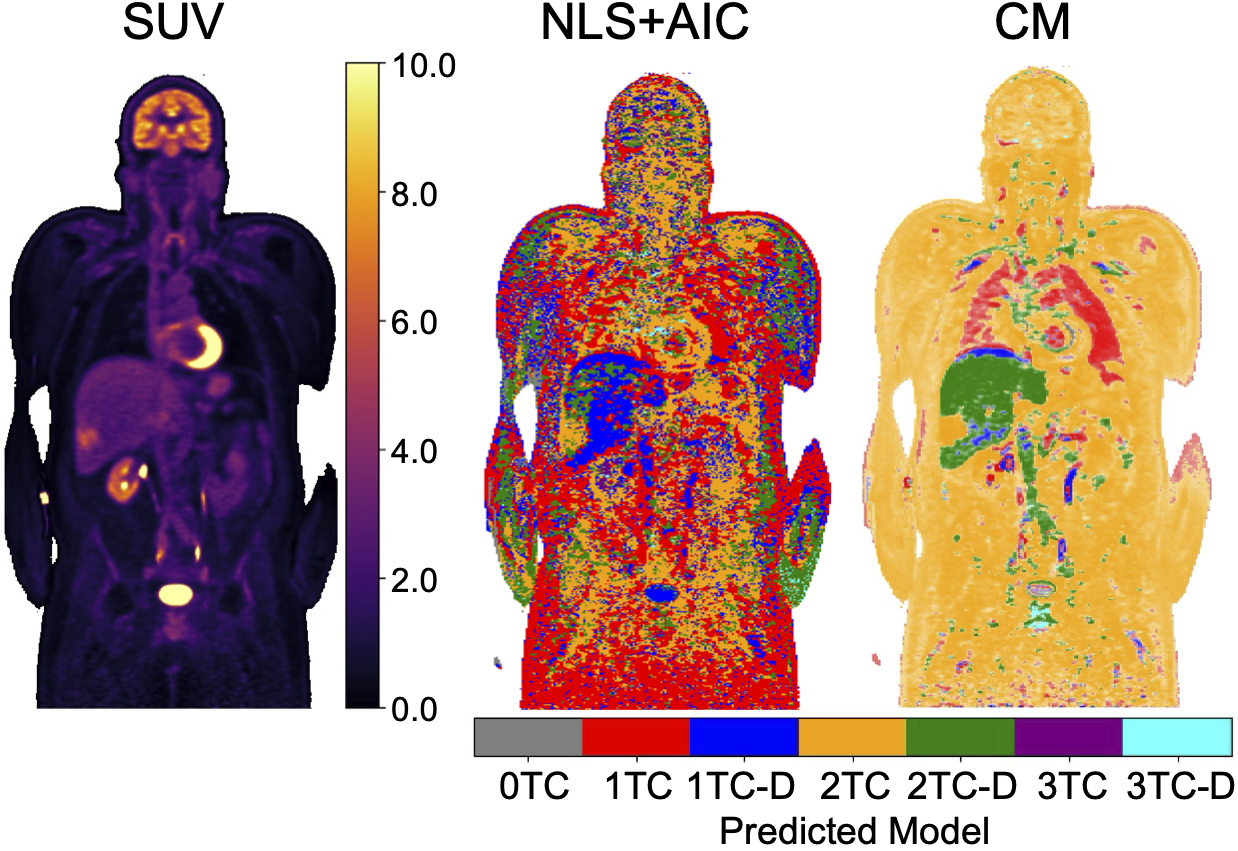}
    \caption{SUV image and voxelwise kinetic model selection maps from NLS + AIC and CM for a coronal total-body FDG PET slice. Colors identify the most probable model, and in CM map opacity represents its model probability (more opaque = higher probability)
    }
    \label{fig: compartment_realdata_compare}
\end{figure}

\begin{figure*}[tb!]
    \centering
    \includegraphics[width=0.85\textwidth]{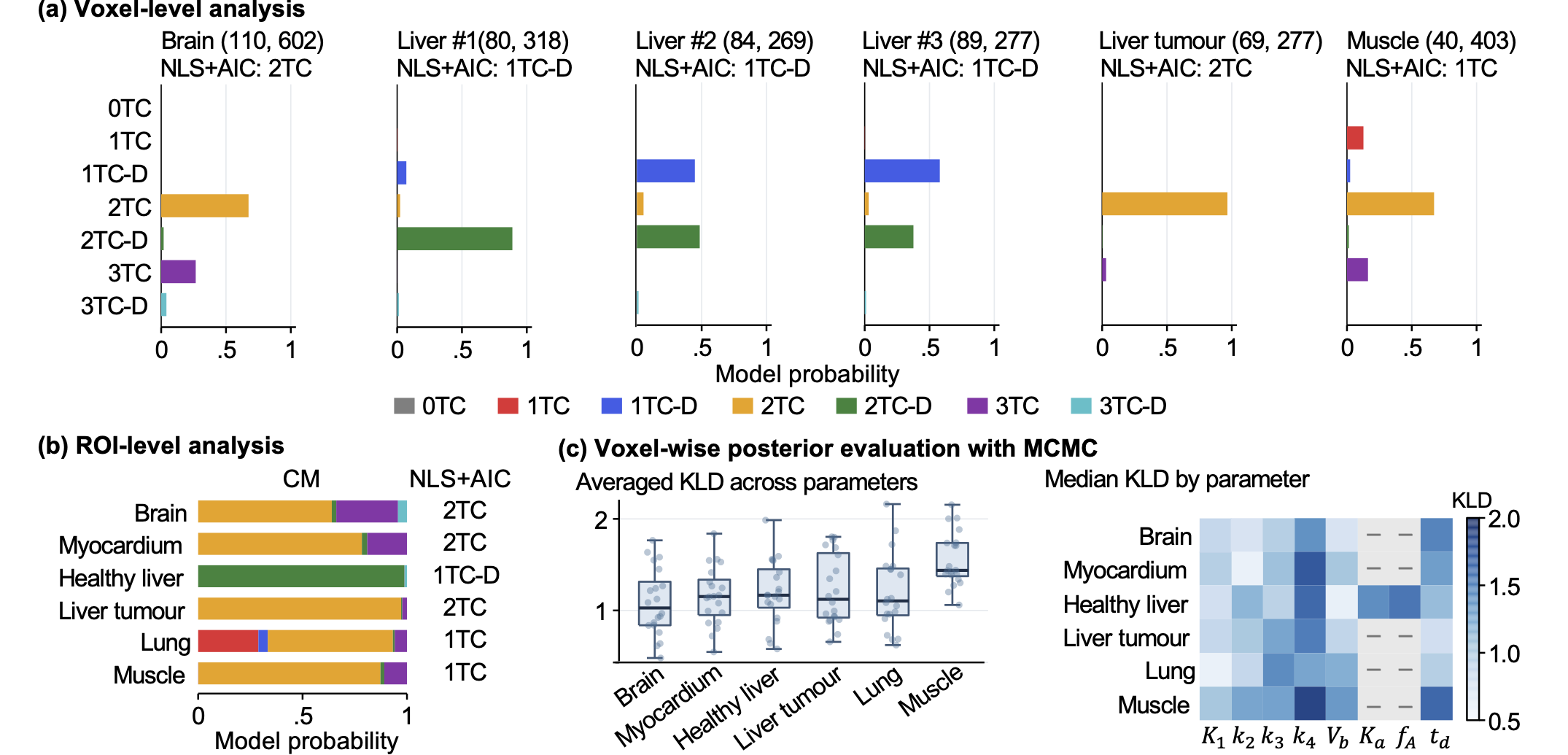}
    \caption{Clinical voxel- and ROI-level model selection and posterior analysis. (a) Model probabilities for selected voxels, with corresponding NLS+AIC model selections. (b) Model probabilities inferred from ROI-averaged TACs and NLS+AIC selections. (c) CM–MCMC posterior agreement: voxel-wise mean KLD across parameters, grouped by ROI, and parameter-wise median KLD within each ROI.
    }
    \label{fig: compartment_P001_analyses}
\end{figure*}

\begin{figure}[tb!]
    \centering
    \includegraphics[width=0.75\columnwidth]{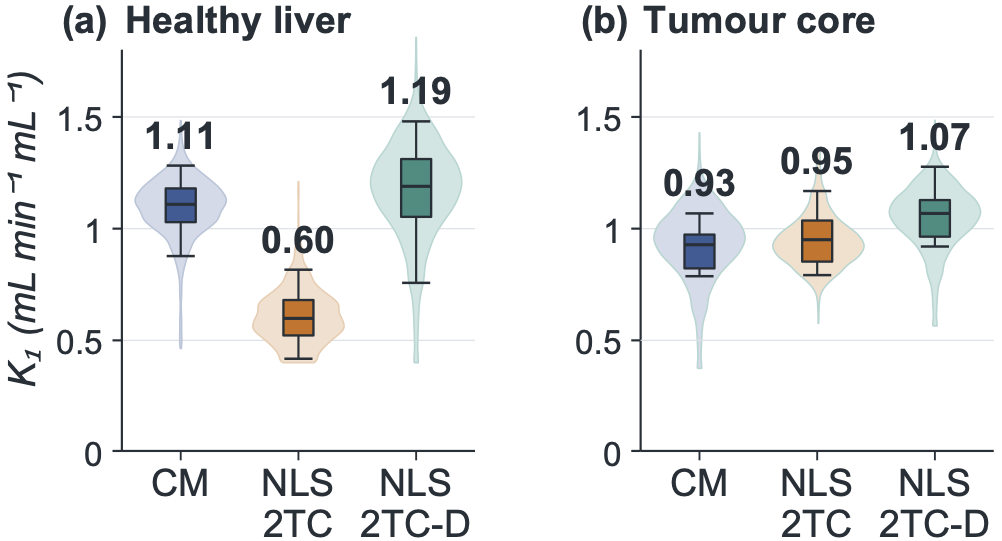}
    \caption{Voxelwise $K_1$ estimates in healthy liver and liver tumour from CM (posterior medians under the selected model) and NLS fits of 2TC and 2TC-D. Boxes show medians and interquartile ranges; pale violins show distributions across voxels.
    }
    \label{fig: Liver_tumour_K1}
\end{figure}

When applied to the clinical FDG TB-PET scan of subject P001 (Fig.\ref{fig: compartment_realdata_compare}), the CM generates an anatomically coherent whole-body model selection map that resolves organ-specific kinetic heterogeneity. 
The NLS+AIC map is very noisy, often selecting the 1TC model in the brain, muscle, and soft tissues, and 1TC-D in the liver. In contrast, the CM map is much smoother and matches the expected organ physiology, selecting 1TC in the lungs, 2TC in the brain, myocardium, skeletal muscle, and kidneys, and 2TC-D across the healthy liver.
Crucially, within the inferior right hepatic lobe, both methods isolate the known neuroendocrine metastasis as a contiguous cluster of single-input 2TC voxels against the surrounding dual-input 2TC-D healthy liver background though NLS+AIC produces 2TC-D voxels in the lesion.

Fig.\ref{fig: compartment_P001_analyses} details this clinical behavior at the individual voxel and regional levels. 
In high-count, homogeneous tissues such as cerebral gray matter (Brain (110, 602)) and the core of the hepatic metastasis (Liver tumour (69, 277)), CM and NLS+AIC both select 2TC. In healthy liver, however, NLS+AIC selects the simpler 1TC-D model, whereas the CM identifies 2TC-D with high confidence (Liver \#1) and quantifies mixed kinetics across 2TC-D, 1TC-D, and 3TC-D in heterogeneous voxels (Liver \#2 and Liver \#3). Similarly, in low-count skeletal muscle (Muscle (40, 403)), NLS+AIC selects 1TC, whereas the CM favors 2TC while assigning secondary probability mass to 3TC and 1TC to reflect low-SNR ambiguity. 
When TACs are averaged within each ROI to reduce noise (Fig. \ref{fig: compartment_P001_analyses}b), the CM chooses 2TC for the brain, myocardium, liver tumor, and skeletal muscle, 2TC-D for the healthy liver, and 1TC for the lung, whereas NLS+AIC selects the simpler 1TC-D model for the healthy liver and 1TC for muscle.
Across voxels sampled per ROI, CM parameter posteriors maintain close agreement with MCMC (median voxel-wise KLD $\approx 0.7$--$1.6$; Fig.\ref{fig: compartment_P001_analyses}c), with the lowest divergence observed for $K_1, V_b$, and $t_d$ across all tissues.

Fig.\ref{fig: Liver_tumour_K1} demonstrates the quantitative bias introduced in clinical parametric imaging when model selection is omitted and an inappropriate kinetic model is used across the liver. 
In healthy liver voxels (Fig. \ref{fig: Liver_tumour_K1}a), the CM selects the 2TC-D model and estimates a median $K_1$ of $1.11~\text{mL}\cdot\text{min}^{-1}\cdot\text{mL}^{-1}$, which closely matches the NLS 2TC-D estimate ($1.19~\text{mL}\cdot\text{min}^{-1}\cdot\text{mL}^{-1}$). In contrast, fitting the single-input 2TC model with NLS ignores the portal venous input and underestimates $K_1$ by nearly half ($0.60~\text{mL}\cdot\text{min}^{-1}\cdot\text{mL}^{-1}$).
Conversely, in the liver tumor core (Fig.\ref{fig: Liver_tumour_K1}b), the CM selects the single-input 2TC model and estimates a median $K_1$ of $0.93~\text{mL}\cdot\text{min}^{-1}\cdot\text{mL}^{-1}$, which closely agrees with the NLS 2TC estimate ($0.95~\text{mL}\cdot\text{min}^{-1}\cdot\text{mL}^{-1}$). In contrast, fitting the dual-input 2TC-D model with NLS overestimates $K_1$ ($1.07~\text{mL}\cdot\text{min}^{-1}\cdot\text{mL}^{-1}$) and increases its variance across voxels.


\subsection{Brain Dopamine Activation Map}
\begin{figure}[tb!]
    \centering
    \includegraphics[width=0.90\columnwidth]{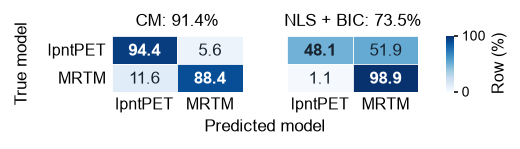}
    \caption{Confusion matrices for MRTM and lpntPET selection on simulated PET data using CM and NLS + BIC. Rows indicate true models and columns predicted models; entries show counts and row-normalized percentages.
    }
    \label{fig: lpnt_confusion_matrix}
\end{figure}

\begin{figure}[tb!]
    \centering
    \includegraphics[width=\columnwidth]{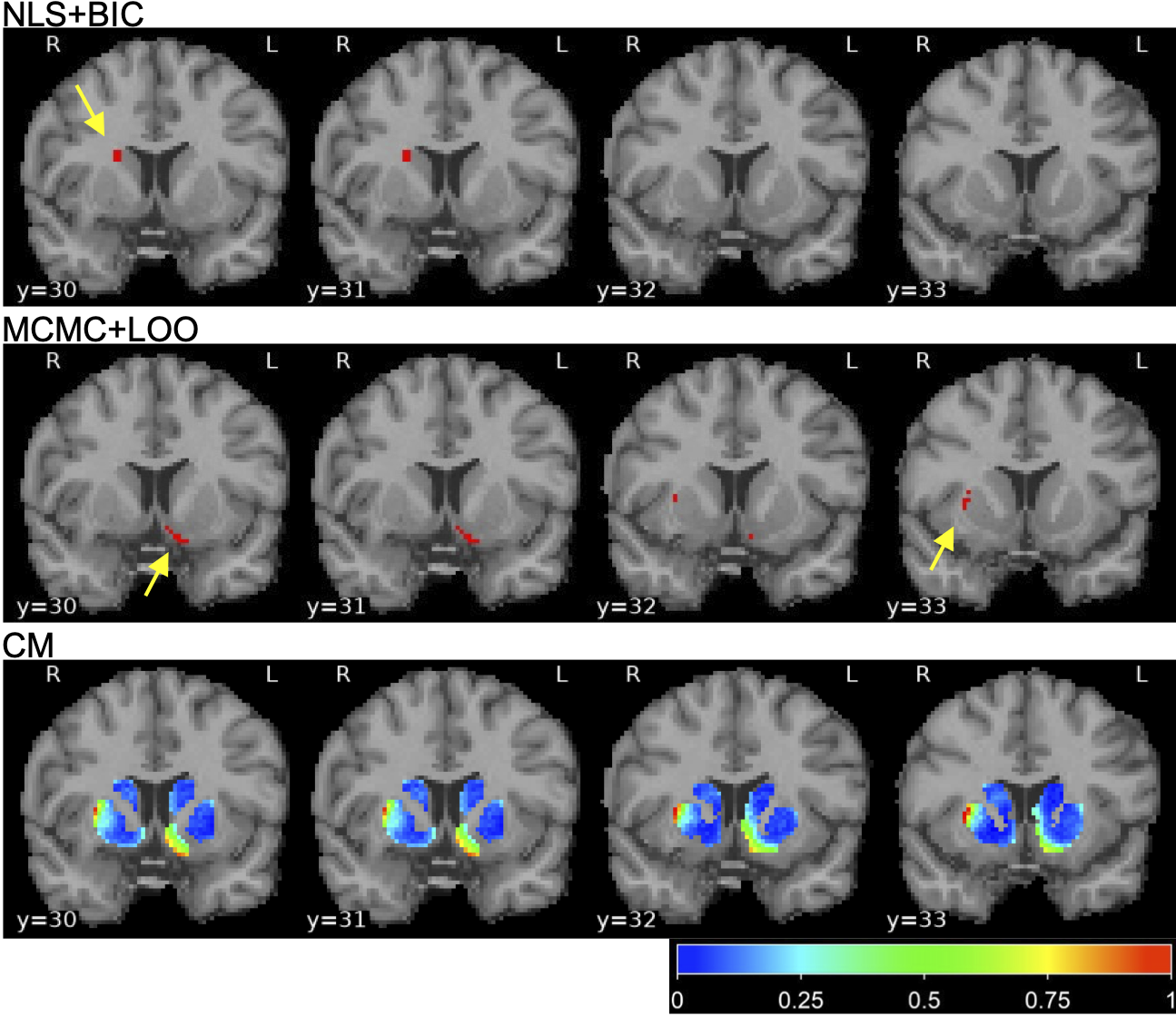}
    \caption{Voxelwise lpntPET model selection in the striatum for real data, overlaid on anatomical images across four coronal slices. NLS+BIC and MCMC+LOO provide binary classifications, with red indicating voxels assigned to lpntPET. CM provides lpntPET model probabilities.
    }
    \label{fig: lpnt_selection_map}
\end{figure}
We compared our CM against NLS paired with the Bayesian Information Criterion (BIC) in this case study, as Normandin \emph{et al.}~\cite{normandin2012linear} demonstrated that BIC provides the most accurate model selection between MRTM and lp-ntPET. Across the 2,000 simulated test TACs (Fig.\ref{fig: lpnt_confusion_matrix}), the CM achieves classification accuracy of 91.4\%, outperforming NLS+BIC at 73.5\%. In particular, the CM maintains balanced sensitivity and specificity, correctly identifying 94.4\% of activated lp-ntPET curves and 88.4\% of baseline MRTM curves. 
In contrast, NLS+BIC achieves high specificity for MRTM (98.9\%) but misses more than half of the true dopamine activation curves, identifying only 48.1\% of lp-ntPET TACs with 51.9\% misclassified as MRTM. 
Furthermore, Table~\ref{tab:posterior_performance} Part~B shows that the CM accurately recovers the kinetic parameter posteriors for both models relative to the reference MCMC distributions. For MRTM, median KLD values range from 0.65 to 0.78 (median APE: 7.6\%--13.9\%). For lp-ntPET, the delivery and efflux parameters ($R_1, k_2, k_{2a}$) and the neurotransmitter response parameters ($\gamma, t_D, t_P, \alpha$) yield median KLD values between 0.55 and 1.32 and median APEs between 7.1\% and 19.8\%.

Fig.\ref{fig: lpnt_selection_map} compares the voxel-wise striatal dopamine activation maps produced by NLS+BIC, MCMC+LOO, and the proposed CM. Consistent with its low sensitivity in the simulation study, NLS+BIC detects almost no smoking-induced dopamine release, identifying only a tiny, isolated cluster of activated voxels in the right putamen. 
Similarly, the binary classification map from MCMC+LOO identifies only sparse, fragmented clusters of activated voxels localized to the left ventral striatum/substantia nigra border and the right putamen. 
In contrast, the CM generates a spatially continuous probability map of dopamine activation ($P(\text{lp-ntPET} | \mathbf{y})$) across the striatum. High activation probabilities ($P > 0.6$, yellow to red) are clearly localized in the bilateral ventral striatum, the lateral right putamen, and focal subregions of the dorsal caudate, whereas surrounding background striatal regions maintain low activation probabilities ($P < 0.25$, blue).


\section{Discussion}
In this work, we developed a CM that performs Bayesian kinetic model selection and parameter estimation simultaneously. 
Traditional frequentist methods (NLS+AIC and NLS+BIC) fit every candidate model separately at each voxel using iterative optimization. Because voxel-level PET curves are noisy, NLS often gets trapped in local minima, and the Gaussian noise assumptions behind AIC and BIC break down under low-count photon statistics~\cite{wang2022total, barrett1994noise}. 
Classical Bayesian methods are limited by their low time efficiency and need for manual tuning.
MCMC+LOO~\cite{irace2020bayesian, vehtari2017practical} requires thousands of sequential leapfrog steps for each candidate model, while PET-ABC~\cite{fan2021pet} relies on selecting a proper summary statistic and calibrating the tolerance threshold.
By learning the joint posterior mapping $p(\boldsymbol{\theta}, m | \mathbf{y})$ during training, the CM shifts this heavy computational cost upfront. At inference time, it processes a coronal slice containing around $84,000$ TACs in under 4 seconds on a GPU workstation with 4 NVIDIA A100 GPU, making Bayesian model selection practical for TB-PET clinical research and imaging.

In the total-body FDG study, our results explain why standard NLS+AIC often selects simple models in noisy voxel-level data. 
When voxel noise is nontrivial, the improvement in curve fitting from adding a third compartment or a dual blood input is masked by random noise fluctuations. 
Consequently, the AIC parameter penalty dominates and forces NLS+AIC to favor simpler models, misclassifying most simulated 3TC and 3TC-D curves as 2TC or 2TC-D (Fig.\ref{fig: compartment_confusion_matrix}) and selecting 1TC in clinical skeletal muscle and 1TC-D in healthy liver (Fig.\ref{fig: compartment_realdata_compare}). 
In contrast, the CM learns how noise affects TACs across all candidate models and retains high selection accuracy across both simple and complex models. 
Clinically, selecting the right model prevents severe errors in parameter estimation (Fig.\ref{fig: Liver_tumour_K1}). 
Healthy liver tissue receives blood from both the hepatic artery and the portal vein, whereas liver metastases are supplied mainly by the hepatic artery~\cite{wang2018dynamic, wang2022total}. By identifying the dual-input 2TC-D model in healthy liver and the single-input 2TC model in the liver tumor, the CM avoids the $\sim 50\%$ underestimation of perfusion ($K_1$) caused by forcing a single-input 2TC model onto normal liver, as well as the upward bias and extra variance caused by applying a dual-input model to the liver tumor.

A similar advantage is seen in the brain $[\text{}^{11}\text{C}]\text{raclopride}$ study. Detecting transient dopamine release requires distinguishing a small, brief dip in the TAC (modeled by the lp-ntPET model) from baseline receptor kinetics (modeled by the MRTM)~\cite{normandin2012linear}. 
Although Normandin \textit{et al.}~\cite{normandin2012linear} showed that BIC controls false positives better than AIC or F-tests for lp-ntPET, its strict complexity penalty ($n_m \ln M$) heavily penalizes the four extra neurotransmitter parameters under voxel-level noise. As a result, NLS+BIC misses more than half of the simulated dopamine curves (48.1\% sensitivity) and fails to detect smoking-induced striatal activation in the clinical scan (Fig.\ref{fig: lpnt_selection_map}), consistent with known sensitivity trade-offs in voxel-wise lp-ntPET~\cite{liu2019controlling, morris2024modeling}. 
While MCMC+LOO improves sensitivity, hard binary thresholding on noisy voxel-wise LOO scores produces scattered, fragmented clusters. 
Because the CM directly outputs continuous posterior activation probabilities $P(m = 1 | \mathbf{y})$, it produces smooth, bilateral activation maps across the ventral striatum (nucleus accumbens, which mediates smoking reward and craving) and dorsal striatum (putamen and caudate, involved in habit formation), matching prior findings in tobacco addiction studies~\cite{zakiniaeiz2022nicotine, martinez2003imaging}.

Several limitations and future directions should be noted. 
First, the CM is trained on simulated TACs generated using a specific frame schedule and arterial input functions of our FDG cohort, so applying it to a different scanner protocol or tracer requires retraining or fine-tuning. In prior work, we showed that lightweight low-rank adaptation (LoRA) can help adapt a pretrained CM to new protocols in minutes\cite{zhao2026general}.
Second, in the current study, we estimated kinetic parameters (such as $K_1$ in Fig.\ref{fig: Liver_tumour_K1}) using only the single ``winning'' model with the highest posterior probability, $\hat{m} = \arg\max_k P(m = k | \mathbf{y})$. 
However, our clinical results (Fig.\ref{fig: compartment_P001_analyses}a) show that many voxels, such as those in the liver and skeletal muscle, do not follow a single pure model, but instead assign non-zero probabilities to multiple models at once. In practice, mixed voxel kinetics arise naturally from partial volume effects, blood or tissue spillover at organ and lesion boundaries, and cellular heterogeneity within tumors~\cite{wang2022total, cunningham1993spectral}. 
Relying only on the single winning model ignores this uncertainty and can create artificial step changes between neighboring voxels when two competing models have similar probabilities. 
In statistical theory, this problem is addressed by Bayesian model averaging (BMA)~\cite{hoeting1999bayesian} and multimodel inference~\cite{burnham2002model, stephan2010ten}. Instead of picking a single winner, BMA computes a shared physiological parameter $\Delta$ (such as perfusion $K_1$) as a weighted average across all candidate models using their posterior probabilities: $p(\Delta | \mathbf{y}) = \sum_{k=0}^{K-1} p(\Delta | m = k, \mathbf{y})\, P(m = k | \mathbf{y})$.
While classical spectral analysis~\cite{cunningham1993spectral} also models mixed tissue components as a linear sum of basis functions, it does not provide posterior probabilities over discrete compartmental hypotheses. Because our CM already draws joint posterior samples $(\boldsymbol{\theta}^{(n)}, m^{(n)}) \sim p(\boldsymbol{\theta}, m | \mathbf{y})$ across all models, computing BMA parameter estimates averages the shared parameter samples across all models. 
In future clinical studies, combining CM posterior model probabilities through BMA could provide more accurate parametric maps across tissue boundaries and help quantify mixed vascular and metabolic underpinnings inside heterogeneous tumors.

\section{Conclusion}
We developed a CM for joint voxel-wise Bayesian kinetic model selection and parameter estimation in dynamic PET. Across [$^{18}\mathrm{F}$]FDG TB-PET and brain [$^{11}\mathrm{C}$]raclopride neurotransmitter imaging, the CM improved model selection accuracy over conventional methods and matched reference MCMC parameter posteriors at a fraction of the computational cost. It produces posterior model probabilities and parameter distributions in seconds per slice, making Bayesian kinetic modeling practical for clinical and research applications. A demonstration of FDG TB-PET model selection is available at \url{https://github.com/yundumbledore/PET-TRACER}.

\section*{Acknowledgment}
The authors acknowledge the facilities and the scientific and technical assistance of Sydney Imaging, a core research facility at The University of Sydney, and the National Imaging Facility, a National Collaborative Research Infrastructure Strategy (NCRIS) Capability.

\bibliographystyle{IEEEtran}
\bibliography{bib}

\end{document}